\documentclass[preprint,preprintnumbers,amsmath,amssymb,elsart]{revtex4}
\usepackage{graphicx,psfrag,bbm,latexsym,color,dcolumn,bm,dsfont,bbm,color,mathrsfs,bbold,latexsym,amsmath,amsfonts,
amssymb,epsfig,natbib}

\newcommand{\beq}{\begin{equation}}
\newcommand{\eeq}{\end{equation}}

\newcommand{\beqa}{\begin{eqnarray}}
\newcommand{\eeqa}{\end{eqnarray}}

\begin{document}
\title{The Quantum Noise of Ferromagnetic $\pi$-Bloch Domain Walls}
\author{P.R. Crompton}
\affiliation{Department of Applied Maths, School of Mathematics, University of Leeds, Leeds, LS2 9JT, UK.}
\vspace{0.2in}
\date{\today}

\begin{abstract}\vspace{0.2in}
{We quantify the probability per unit Euclidean-time of reversing the magnetization of a $\pi$-Bloch vector, which describes the Ferromagnetic Domain Walls of a Ferromagnetic Nanowire at finite-temperatures. Our approach, based on Langer's Theory, treats the double sine-Gordon model that defines the $\pi$-Bloch vectors via a procedure of nonperturbative renormalization, and uses importance sampling methods to minimise the free energy of the system and identify the saddlepoint solution corresponding to the reversal probability. We identify that whilst the general solution for the free energy minima cannot be expressed in closed form, we can obtain a closed expression for the saddlepoint by maximizing the entanglement entropy of the system as a polynomial ring. We use this approach to quantify the geometric and non-geometric contributions to the entanglement entropy of the Ferromagnetic Nanowire, defined between entangled Ferromagnetic Domain Walls, and evaluate the Euclidean-time dependence of the domain wall width and angular momentum transfer at the domain walls, which has been recently proposed as a mechanism for Quantum Memory Storage.}
\end{abstract}

\maketitle

\section{Introduction}
Spin Polarised Scanning Tunnelling Microscopy (SP-STM) measurements of the differential conductance profile of Ferromagnetic Nanowires, such as samples of Fe on W(110) \cite{1}\cite{4}, highlight the importance of understanding the effective polarization of the surface-probing tip subsystem, and its contribution to the measured anisotropy of a Ferromagnetic surface. This is because the response of the surface to the magnetic field of the applied probing tip is dependent on the local projection of the surface magnetization onto the probing tip, via the torque that is induced by the spin-polarized tunneling current. Hence, this mechanism is being actively considered for practical Quantum Memory Storage, via the transfer of angular momentum into Ferromagnetic Domain Walls \cite{14}\cite{13}\cite{16}\cite{15}. Although the corresponding changes in the measured resistivity are not large, it is relatively straight forward to understand the scattering mechanism induced by this torque in the two limits that the domain walls are either very narrow or very broad \cite{10}\cite{11}. In the first case, spin magnons will be specularly reflected without the transfer of angular momentum, whilst in the second case, spins adiabatically realign around the local magnetization, whilst precessing with a characteristic Larmor frequency, and transfer angular momentum into the domain wall. A simple (but effective) picture of the local projection of the spins can be parameterized by defining a Bloch vector between neighbouring Ferromagnetic Domain ($\pi$-Bloch) Walls, 
\beq
{\rm{sin}}( \varphi(x) ) = {\rm{tanh}}[\,(x-x_{0})\, /\, (w/2)\, ]\,,
\eeq

where $x_{0}$ is the position of the first domain wall, and $w$ is the domain wall width. This relation is obtained by assuming that nonlocal surface effects can be neglected, and by minimising the free energy, where the free energy of this system is expressed via the generic double sine-Gordon model,

\beq
S = \int \! dx \,\,  A\left( \frac{\partial \varphi}{\partial x}\right)^{2} + M B \, {\rm{cos}} \varphi  + K \, {\rm{sin}}^2 \varphi\,,
\eeq

where $A$ is the exchange interaction stiffness, $K$ the effective anisotropy of the surface, $M$ the saturation value of the local magnetization vector, and $B$ the magnitude of the applied field which generates Zeeman splitting \cite{8}. The domain wall width, $w$, is then obtained explicitly by assuming that the dissipative dynamics of the spins obeys the Landau-Lifshitz-Gilbert equations, giving,

\beq
w = \,\,\sqrt{ \frac{A}{ K + MB/2 } } \,\,\,\, {\rm{sinh}}^{-1} \!\!\left(\!\sqrt{ \frac{2K}{MB}} \,\right)\!.
\eeq

However, whilst several experiments have confirmed that this locally planar magnon picture is essentially correct \cite{1}\cite{4}\cite{16}\cite{15}, this does mean that the domain wall width of the Ferromagnetic Nanowire is parameterised in (3) via the non-universal effective anisotropy value $K$. This parameterisation therefore limits the accurate calibration of the angular momentum transfer of a Ferromagnetic Nanowire for a practical Quantum Memory Storage device, and more detailed treatments have stressed the importance of also considering noncollinear magnetization effects \cite{14}\cite{sp-stm_qd_theory}.

To renormalize this effective anisotropy value, $K$, we now treat the more general problem of minimising nonadiabatic dynamics in (2), and consider the entanglement between Ferromagnetic Domain Walls that is induced by rapid changes in the tunneling current. We perform a path integral over Euclidean-time for (2) to do this, which is made equivalent to Imaginary-time in our formalism via an exact Wick-rotation \cite{me2}\cite{me}\cite{me4}. Physically, this treatment corresponds to the situation where if the system is heated, Stoner excitations eventually destroy the ferromagnetic order by overcoming local energy barriers, however, rapid recooling is not then guaranteed restore the system to the same initial local valley states and so the free energy minima of (2) is not unique \cite{18}. A similar approach to the one we present has been developed in \cite{18}\cite{19}\cite{20} to treat the Euclidean-time dependence of the nucleation of the Ferromagnetic Domain Walls by applying Langer's Theory. The approach is to find the minima of the free energy in (2), parameterized as a dilute gas of soliton and anti-soliton states, and to determine the effect of small perturbations to this system obtained by adding a single soliton to the grand canonical partition function of the system . This gives a probability per unit time of reversing the magnetization of the Bloch vector described by the Van't Hoff-Arrehenius law,

\beq
\Gamma = \Omega e^{-\beta E_{s}}
\eeq
\beq
\Omega = \kappa \sqrt{\frac{2\beta}{\pi^{3}}} \sqrt{E_{s}} L \sqrt{ \frac{ \prod_{i=1} \Lambda_{i}^{(0)} }{ |\Lambda_{1}^{(\phi)}| \prod_{j>1} \Lambda_{j}^{(\phi)} }} \sqrt{ \frac{ \prod_{i} \Lambda_{i}^{(0)} }{ \prod_{j>1} \Lambda_{j}^{(p)} }}
\eeq

where $E_{s}$ is the saddlepoint solution for the free energy, $\beta$ is Euclidean-time, $\kappa$ is the growth rate of unstable deviations, $L$ is the system length, and $\{\Lambda\}$ are the eigenvalues of the Hamiltonians describing the zero mode, $(0)$, in-plane, $(\phi)$, and out-of-plane fluctuations, $(p)$, of the Ferromagnetic Domain Wall Bloch vectors \cite{18}. In order to treat perturbations of arbitrary size, in our approach, however, we formulate a fugacity expansion for the full grand canonical partition function, which consists of $N$ soliton and anti-soliton states \cite{me5}\cite{me3}. Minimizing the free energy of this full partition function enables us to normalize the probability per unit time in (4) via an additional nonanalytic factor, which corresponds to the entropy of adding an additional $(N-1)$ solitons into the system. We then calibrate this contribution as a phase factor, via the Wick-rotation of the operators of our formalism \cite{me2}\cite{me}\cite{me4}, and use this value to quantify the change in the physical scale of the Bloch vector in (1) and (3) due to the nonanalytic entanglement contributions (quantum noise) arising from magnetization-reversal tunneling transitions between different Ferromagnetic Domains across the surface of the Nanowire.

\section{Mixed Quantum Spin Chain}
Following \cite{me5}, to represent the double sine-Gordon model in (2), we treat the model which consists of an antiferromagnetic (AFM) periodic mixed-spin chain of the form, $1-1-3/2-3/2$, where $1$ and $3/2$ represent the spin magnitudes, and the period of the alternation is four lattice sites. The Hamiltonian of this model is given by,

\beq
H = \sum_{j=0}^{N/4-1} J_{1 \,, 1} {\bm S}^{(1)}_{4j} . {\bm S}^{(1)}_{4j+1} +
J_{1 \,, 3/2} {\bm S}^{(1)}_{4j+1} . {\bm S}^{(3/2)}_{4j+2}  +
J_{3/2 \, , 3/2} {\bm S}^{(3/2)}_{4j+2} . {\bm S}^{(3/2)}_{4j+3} +
J_{3/2 \,, 1} {\bm S}^{(3/2)}_{4j+3} . {\bm S}^{(1)}_{4j+4}
\eeq

where $j$ is the lattice site index of the cell along the spin chain, $J_{a, b}$ is the nearest neighbour spin interaction coupling between the neighbouring spins of magnitude $a$ and $b$, and ${\bm S}^{(a)}$ is the spin operator of magnitude $a$. We keep the ratio between like-like and like-dislike spin vectors fixed such that $J_{1 \,, 1} = J_{3/2 \,, 3/2} = J$ and $J_{1 \,, 3/2} = J_{3/2 \,, 1} = \alpha J$. The phase space of the general quantum spin chain is related to the unstable renormalization group flow of the $SU(2)$ Wess-Zumino-Witten model and the emergence of a gapless groundstate in half-integer spin chains - whilst gapped groundstates states are realised by quantum spin chains with integer spin values \cite{mus1}\cite{wil}. These states can be identified in the above Hamiltonian by constructing the symmetric Young tableaux of the spin-$3/2$ operator (which is an $SU(2)$-spinor) following Haldane's arguments for the destructive interference of quantum fluctuations for antisymmetric states \cite{young}. This analysis reveals that there are three topologically distinct realizations of the gapped groundstate of the above mixed-spin model. The importance of this particular model is, therefore, that it undergoes two topological transitions at zero temperature, where the system is driven through the vacuum angle $\theta=\pi$, by varying the coupling ratio $\alpha$. We therefore define the Bloch vectors of this system, not with respect to the operator algebra defined between local spin sites, $i$, but with respect to a path integral over Euclidean-time given by the grand canonical partition function, $\mathcal{Z}$, defined by summing over all values of the vacuum angle $\theta$,

\beq
\mathcal{Z} = \int \mathcal{D}\theta  \,\,\, {\rm{exp}}\!\left[\,\int_{0}^{\beta} d\tau H \,\, \,\right]  \equiv \int \mathcal{D}\theta  \,\,\, {\rm{exp}}\!\left[\,\int_{0}^{\beta} d\tau
\,\, A_{\bm{n}}\,i\phi-V_{\bm{n}} \,\, \right]
\eeq

\beq
 A_{\bm{n}} \equiv \sum_{i,\,\tau}^{N \otimes T} \,\, \sum_{{\bm S}} \lambda^{i\tau{\bm S}}_{\bm{n}}
\frac{\langle \bm{n}\oplus \bm{1}_{i{\bm S}} \oplus \bm{1}_{\tau{\bm S}}| \theta \rangle}{\langle \bm{n}| \theta \rangle}\,, \quad
V_{\bm{n}} \equiv \sum_{i,\,\tau}^{N \otimes T} \,\, \sum_{{\bm S}}(\lambda^{i\tau{\bm S}}_{\bm{n}})' \frac{\langle
\bm{n}\oplus \bm{1}_{i{\bm S}} \oplus \bm{1}_{\tau{\bm S}}| \bm{n} \rangle} {\langle \bm{n}| \bm{n} \rangle}\,
\eeq

where $A_{\bm{n}}$ and $V_{\bm{n}}$ are defined as the compact and noncompact portions of the spin operators in (6) determined through the nonperturbative matrix elements $\lambda^{i\tau{\bm S}}_{\bm{n}}$ and $(\lambda^{i\tau{\bm S}}_{\bm{n}})'$, and $i$ and $\tau$ are, respectively, space and Euclidean-time indices \cite{me2}\cite{me}. Consequently, there are three compact Bloch vectors defined in the regions between the zero temperature transitions of the above system, which we use to represent a Ferromagnetic Nanowire consisting of four anti-aligned $\pi$-Bloch domain walls. The above lattice operator formalism is explicitly related to the effective action of the double sine-Gordon model in (2) via a two-loop expansion of the following form,

\beq
S_{eff} = \int_{0}^{\beta} \!\! \int_{-\pi}^{\pi} \!\! d^{2}x \,\, a_{\tau}\,\partial^{2}_{x} A_{\bm{n}} - a_{i}\,\partial^{2}_{x} V_{\bm{n}}   -
a_{i}\wedge a_{\tau}
\eeq

where the Euclidean-time and spatial lattice spacings are quantified via,
\beq
\theta \, a_{i} \equiv A_{\bm{n}(i)} -\langle A_{\bm{n}} \rangle_{N,T}\,, \quad
\beta  \, a_{\tau} \equiv A_{\bm{n}(\tau)} - \langle A_{\bm{n}} \rangle_{N,T}
\eeq

which quantify the out-of-plane and in-plane fluctuations of the Ferromagnetic Domain Walls (Bloch vectors), respectively.

As indicated by the operator formalism in (9) and (10), since strictly no measurement of the inverse spin gap of a finite-size quantum spin chain system can be a divergent function, the topological transitions (and hence the Bloch vectors) of the lattice model in (7) and (8) are only defined upto the intrinsic lattice IR cutoff scale given by the nonperturbative dynamics of $\lambda^{i\tau{\bm S}}_{\bm{n}}$ and $(\lambda^{i\tau{\bm S}}_{\bm{n}})'$. There is, therefore, a nonanalytic first order region around the topological transition value $\theta=\pi$ in the quantum spin chain at finite temperatures, which represents the mismatch in the overlap of $\phi$ and $\theta$ \cite{me2}\cite{me}. Therefore, by finding the maxima of (7), we can treat the entanglement of the Ferromagnetic Domain Walls at finite-temperatures by quantifying these nonanalytic regions, which correspond to a nonadiabatic change in the normalization of $\phi$, from (7) and (8). A similar treatment for renormalizing the double sine-Gordon model describing the Ferromagnetic Domain Walls was given in \cite{18}, by treating these nonanalytic regions using a perturbative expansion scheme.

\section{Renormalized Entanglement Entropy}
We minimize (7), and define an analogous saddlepoint solution to the grand canonical partition function treatment in (5), by first evaluating the path integral in (7) using the conventional continuous-time Quantum Monte Carlo scheme. Practical details of this approach are given in \cite{me5}, and this procedure is used to generate a set of the nonperturbative matrix elements $\lambda^{i\tau{\bm S}}_{\bm{n}}$ and $(\lambda^{i\tau{\bm S}}_{\bm{n}})'$ in (8) for the model defined in (6). The important quantities for determining the saddlepoint solution in (5), are the eigenvalues, $\{\Lambda^{(0)}\}$, $\{\Lambda^{(\phi)}\}$ and $\{\Lambda^{(p)}\}$, which describe the zero mode, in-plane and out-of-plane fluctuations of the Bloch vectors. However, from (10), all of these contributions are incorporated into our nonperturbative matrix elements $\lambda^{i\tau{\bm S}}_{\bm{n}}$ and $(\lambda^{i\tau{\bm S}}_{\bm{n}})'$. Therefore, to quantify this saddlepoint solution in our new approach we explicitly evaluate the determinant of the operator $P$, which equivalently defines the transfer matrix in the method we use for the importance sampling of the path integral,

\beq
{\rm {det}} P = \prod_{i=1}^{N} \Lambda_{i}\,, \quad
\mathcal{Z} = \int \mathcal{D}\theta  \,\,\, P
\eeq

where $\Lambda_{i}$ are the eigenvalues of $P$. The practical reason why only the addition of one extra soliton is considered in (5), is that, in general, the support of the eigenvalues in (11) cannot be expressed in closed analytic form. However, it is possible to locally regularize (11) via the $\zeta$-function renormalization prescription \cite{me4}. Following this prescription, the logarithm of ${\rm {det}} P$ in the limit $N\rightarrow\infty$ can be defined by analytically continuing the following function,

\beq
\sum_{i=1}^{N} \Lambda_{i}^{-s} \, {\rm{ln}}\, \Lambda_{i},
\eeq

from large $s$ to $s=0$. Introducing the $\zeta$-function, $\zeta_{P}(s) = \sum_{i=1}^{N} \Lambda_{i}^{-s}$, the logarithm of the determinant of $P$ is then completely defined by the first derivative of this $\zeta$-function at $s=0$. Writing this $\zeta$-function in the usual form of a Mellin transform,

\beqa
\zeta_{P}(s) &  = & \frac{1}{\Gamma(s)} \int_{0}^{\infty} \, {\rm{Tr}} \, (e^{-\theta P}) \, \theta^{s-1} \, d\theta\\
& \equiv & \int_{0}^{\infty} \, \Lambda(\theta) \,\, \theta^{s-1}  \,\, d\theta
\eeqa
\beq
\label{entropy}
{\rm{ln}}\, {\rm{det}} P  =  -\int^{\infty}_{0} \Lambda(\theta) \,\, {\rm{ln}}(\theta) \,\, d\theta
\eeq
However, although (15) is a well-defined expression for the path-integral in (7), and the local limit of the free energy, because of the rotational symmetry of the Bloch vectors in (8) this minima is not unique, and there are several different realizations of the system in (7) which are degenerate. %However, because of the mismatch in the overlap between $\phi$ and $\theta$ in (7), there is no simple symmetry mapping to describe the relationship between these degenerate minima, and the saddlepoint solution is defined over a range of values of $\phi$. Experimentally this situation is observed when the variation in the average surface energy of a sample is greater than the domain wall width.
For the general solution of (12), not only can the local minima in (15) be finite, but also a number of the higher moments corresponding to the tunneling from two-, three- and (and more) soliton states. Therefore, to determine the saddlepoint solution of (7), we have to evaluate the $\zeta$-function prescription at a different point on the fundamental strip, at $s=1$, for each of these moments,

\beq
\frac{ d^{j} \zeta_{P}(s) }{ ds^{j} } \mid_{s=1}  =  \int_{0}^{\infty} \,\, \Lambda(\theta) \,\, {\rm {ln}}^{j}(\theta) d \theta, \quad j = 0 ...N
\eeq

which allows us to address the choices of branch for all of these higher moments that relates to the degeneracy of the minima in (15). Collectively minimising these moments, at $s=1$, then becomes equivalent to analytically continuing the following function via the $\zeta$-function prescription in (12),

\beq
H = \int_{0}^{\infty} \,\, \Lambda(\theta) \,\, {\rm {ln}}\Lambda(\theta) \,\, d \theta
\eeq

which corresponds to the familiar Von Neumann form of the entropy of a quantum spin system. To find the saddlepoint solution of the full grand canonical partition function in (7), it is therefore analytically more well-defined to maximize the entanglement entropy in (17), rather than minimize the free energy in (15), since this resolves the issue of the degeneracy of the free-energy minima. However, it follows from the $\zeta$-function prescription that the local part of the entanglement entropy in (17) is then, in general, a divergent function. Formally, the structure of this singularity is a polynomial ring. 

Having maximized the path integral in (7) via importance sampling, we plot the quantity ${\rm {ln}}\,\Lambda_{i}$ as a function of $\phi$, from our definitions in (7) and (11), in Figure 1. This quantity defines the normalized density of the logarithm of the eigenvalues along the fundamental strip, at $s=1$, and corresponds to the entanglement entropy in (17), for the discrete system in (7). A more detailed discussion of the system-size of this treatment is presented in \cite{me5}, but the general trends evident in Figure 1 (for $N=256$) are that there is both a $\phi$-independent component to ${\rm {ln}}\,\Lambda_{i}$, which is relatively flat and centred on ${\rm {ln}}\,\Lambda_{i} \sim 1.9$, and two $\phi$-dependent components to ${\rm {ln}}\,\Lambda_{i}$, which are asymptotic to $\phi \sim 0$ and $\phi \sim 0.9$.
From (17) these contributions correspond, respectively, to the resolution of the branches of the tunneling processes from multiple soliton states at $s=1$ (flat plateau), and the locally-divergent simple spin-wave magnon component defined in (15) (asymptotes). Via the normalization procedure for $\phi$ in (8) we can, therefore, use the above numerical value of the flat plateau to quantify the nonanalytic contribution to the Bloch vectors due to entanglement between the Ferromagnetic Domain Walls, in the Ferromagnetic Nanowire system described by three compact Bloch vectors defined through (6) and (9) by summing over all possible tunnelling processes, via importance sampling.

\begin{figure}
\epsfxsize=3.3 in\centerline{\epsffile{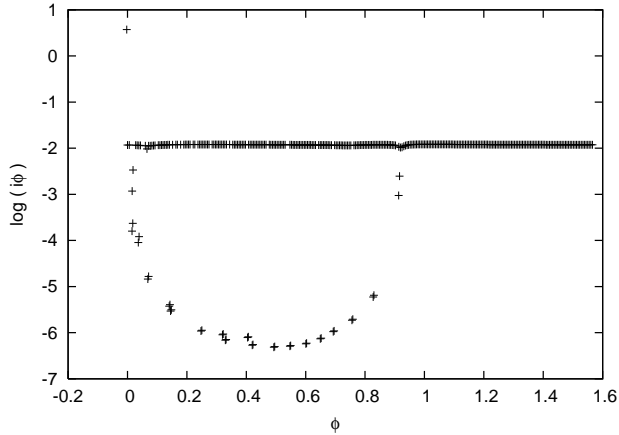}}
\caption{The density of the logarithm of the eigenvalues of the transfer matrix in (11), ${\rm {ln}}\,\Lambda_{i}$, plotted as a function of the Bloch vector angle $\phi$, defined in (8).}
\end{figure}
\section{Entangled Domain Wall Width}

%By defining a renormalization procedure for the quantum spin chain, as we have described in detail in \cite{me2}\cite{me}, we have found a relation in (2) that defines the conformal scaling of the parameters of the double Sine-Gordon model in (1).  We now show that the new relation we have obtained in (2) can be used to define a simple rescaling program for this experimental analysis, in order to quantify the effect of dynamical fluctuations on the ansatz that is used for the differential conductance profile of ferromagnetic domain walls in SP-STM experiments \cite{4}, and to rescale the fitted anisotropy values.

When the double sine-Gordon model is used to describe the magnons associated with the differential conductance profile of ferromagnetic nanowires in \cite{4}\cite{8}, or more formally when this model is used (equivalently) to describe the (infrared) singularities associated with topological changes in the ground states of quantum spin chains (where these domain walls are viewed as solitons) \cite{mus1}\cite{ent}, in both cases, $\phi$, is used to parameterise the angle that describes the orientation of magnetization vectors as they cross the domain walls. The two interaction terms in the double sine-Gordon model in (2) therefore have specific generic meanings - the first interaction term describes the singlet state of this magnetization vector, whilst the second interaction term describes the triplet state found by the Zeeman splitting of this magnetization vector. Crucially, when long-ranged dynamical fluctuations are neglected (in the limits of either very narrow or very broad domain walls), the only physical process that can generate this Zeeman splitting is an applied external field. However, when these long-ranged dynamical fluctuations are included, this splitting can also be generated dynamically by inter-valley tunnelling processes. Thus, to go beyond a simple magnon picture for the model requires a treatment of the (nonlocal) dependence of the couplings of the two interaction terms, in (2), on the long-ranged dynamical cutoff scale of these inter-valley interactions. 

This long-ranged dynamical cutoff scale in our analysis corresponds to relative scale set by the flat plateau in Figure 1, and is defined in normalised units of Euclidean time. By explicit construction of the operators in (8) this cutoff scale is also defined as a function of normalised units of $\phi$. We can infer from our working in the previous section that this long-ranged dynamical cutoff scale is at most logarithmically divergent upto the scale of the magnon (which also follows directly from the Mermin-Wagner theorem on the absence of spontaneous symmetry breaking in $d<2$ \cite{me2}). Therefore, in (2), the coupling of the first interaction term $MB$ can be smoothly deformed into the couplings of the second interaction term $K$, and the elementary magnons of our analysis our defined by the scale upto which they can pass through the domain wall without becoming entangled through inter-valley tunnelling. Since $\phi$ in (8) traces out the surface of a sphere in Euclidean time, this smooth deformation maps this sphere into an ellipsoid. The eccentricity of this ellipsoid, $e$, is defined from (2) by,

\beq
e = \sqrt{ 1 - (\eta/2\gamma + 1)^2 }\, .
\eeq

where $\gamma = K$ and $\eta = MB/2B_{max}$ and the entanglement entropy is defined in units of  $\ln{\rm{B}} = \ln \left(B/B_{max}\right)$. As the elementary magnons of this analysis pass through the domain wall they therefore become effectively oblate through entanglement. In the limit that the triplet interaction term is larger than the (magnon) singlet interaction term in (2), the above eccentricity parameter then becomes purely imaginary - corresponding to a completely entangled state. A similar elliptical parameterisation of the Bloch vectors is given in \cite{17}, to quantify the Pauli cloning of quantum bits.

\begin{figure}
\epsfxsize=3.3 in\centerline{\epsffile{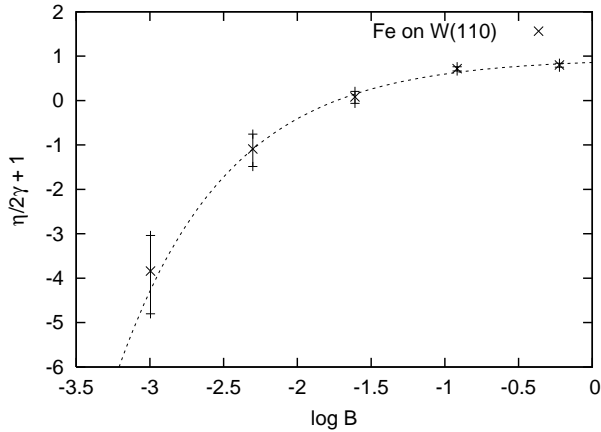}}
\caption{The Lorentz scaling of the effective anisotropy, $K$, in (1) defined as a function of the logarithm of the applied field $B$ using the experimental Ferromagnetic Domain Wall data from \cite{4}. The curve indicates the zero  entanglement boundary of magnons passing through the domain wall, and is obtained from the preceding numerical analysis.}
\end{figure}

In Figure 2 we plot $\eta/2\gamma + 1$ versus $ {\rm{log}} \, {\rm{B}}$ from the domain wall differential conductance profile data obtained experimentally in \cite{4}. To rescale the effective anisotropy data values (which were obtained from fits of the experimental data to the ansatz in (3)) we follow the relatively simple prescription of rescaling $e$, via the graph. We firstly substitute the data values of $\eta$ and $\gamma$ into (18) to calculate the eccentricity, $e$, of the corresponding Bloch vector. We then recalibrate the magnon couplings by defining the new effective anisotropy value to be $\gamma' = e$, and set the eccentricity of this new Bloch vector to unity by reading the corresponding $\eta'$ off the graph. Finally, we calculate the rescaled eccentricity value $e'$, by putting $\gamma'$ and $\eta'$ in (18). Thus, the ratio $e'/e$ quantifies  the entanglement contribution to the effective anisotropy value - as distinct from the Zeeman splitting. From this (renormalization) analysis, the effective anisotropy value differs from the renormalized effective anisotropy value by 11\% at 200mT, which implies that the level of quantum noise and angular momentum transfer associated with the domain walls is 11\% at 200mT. However, this contribution increases dramatically as the field strength is decreased, and magnetic ordering is replaced by quantum-mechanical mixing. The fitted line in Figure 2 corresponds to the scale-independent numerical value for $e$ we obtained from our numerics in the preceding section, which shows good agreement with the data.

%\section{SUMMARY}
\section{Summary}

In this article, we have derived a nonperturbative scaling relation (18) for the angular momentum transfer of magnons at ferromagnetic nanowire domain walls from a rigorous treatment of the singularities associated with the grand canonical partition function of interacting Ferromagnetic and Anti-Ferromagnetic Domain Walls. Our scaling relation in (18) is defined by finding the infrared cutoff dependence of the couplings in the leading order magnon description of changes in resistivity of the domain wall \cite{me5}, and we have used a $\zeta$-function renormalization prescription to parameterise the inter-valley tunneling contributions to the magnon couplings via the construction of an exact polynomial ring \cite{me4}. The advantage of this approach, over existing calculations for the grand canonical partition function of interacting Ferromagnetic and Anti-Ferromagnetic Domain Walls \cite{18}\cite{19}\cite{20}, is that the residual degeneracy of the couplings in this polynomial ring means that we do not need to mathematically separate the planar and non-planar contributions to the domain wall in order to define the magnons, (8). Hence, we are able to quantify the relative changes in quantum noise and angular momentum transfer at the domain wall due to the (inter-valley) mixing of the magnons.  

Although no approximations are made in this nonperturbative renormalization calculation, our analysis rests on the simple magnon picture that is used to parameterise the differential conductance profile at the domain wall. We have good evidence in Figure 1 to believe that the leading order contribution to the saddlepoint solution of our system is at most logarithmically divergent, which is in agreement with the general assumptions of the Mermin-Wagner theorem \cite{me}\cite{me2} and the experimental successes of the magnon approach \cite{14}\cite{15}\cite{10}. However, although we have explicitly evaluated a grand canonical canonical partition function for the interacting Ferromagnetic and Anti-Ferromagnetic Domain Walls consisting of $256$ interacting pairs using importance sampling, the topology of the soliton model we have used only contains interaction terms upto third order. Hence, the numerical results we have presented may not be consistent with much longer ferromagnetic nanowires, consisting of a larger number of domains. Nevertheless, our numerical predictions do show good agreement with the scaling of the domain wall data in \cite{4}, and our general approach is consistent with the experimental findings of the leading order magnon behaviour \cite{14}\cite{16}\cite{15}\cite{11}\cite{10}, which we are now able to renormalize nonperturbatively.

\end{document}